\documentclass[a4paper]{jpconf}
\usepackage{graphicx}
\usepackage{iopams}
\usepackage{amssymb,latexsym}
\usepackage{amstext,amsthm}
\usepackage{amscd}
\usepackage{amsgen,amsfonts,amsbsy}

\theoremstyle{plain}

\theoremstyle{definition}

\newcommand{\Z}{\mathbb{Z}}
\newcommand{\N}{\mathbb{N}}

\newcommand{\C}{\mathbb{C}}
\newcommand{\set}[2]{\{#1|\ #2\}}
\newcommand{\sub}{\subseteq}

\newcommand{\M}{\mathrm{M}}
\newcommand{\GL}{\mathrm{GL}}
\newcommand{\Ad}{\mathrm{Ad}}
\newcommand{\Inn}{\mathrm{Int}}

\newcommand{\diag}{\mathrm{diag}}
 \newcommand{\ket}[1]{|{#1}\rangle}

\newcommand{\SL}{\mathrm{SL}}
\newcommand{\Sp}{\mathrm{Sp}}

\renewcommand{\sl}{\mathrm{sl}}
\renewcommand{\P}{\mathcal{P}}
\renewcommand{\H}{\mathcal{H}}

\begin{document}

\title[Symmetries of finite Heisenberg groups]{Symmetries
of finite Heisenberg groups for $k$-partite systems}

\author{M. Korbel\'{a}\v{r} $^1$ and J Tolar $^2$}
\address{$^1$ Department of Mathematics and Statistics \\
 Faculty of Science, Masaryk University \\ Kotl\'{a}\v{r}sk\'{a} 2, 611 37 Brno,
 Czech Republic \\
 $^2$ Department of Physics \\
 Faculty of Nuclear Sciences and  Physical  Engineering\\
 Czech Technical University in Prague\\ B\v rehov\'a 7,
 115 19 Prague 1, Czech Republic}
\eads{\mailto{miroslav.korbelar@gmail.com},
\mailto{jiri.tolar@fjfi.cvut.cz}}

 \begin{abstract}
Symmetries of finite Heisenberg groups represent an important tool
for the study of deeper structure of finite-dimensional quantum
mechanics. This short contribution presents extension of previous
investigations to composite quantum systems comprised of $k$
subsystems which are described with position and momentum variables
in $\Z_{n_{i}}$, $i=1,\dots,k$. Their Hilbert spaces are given by
$k$-fold tensor products of Hilbert spaces of dimensions
$n_{1},\dots,n_{k}$. Symmetry group of the corresponding finite
Heisenberg group is given by the quotient group of a certain
normalizer. We provide the description of the symmetry groups for
arbitrary multipartite cases. The new class of symmetry groups
represents very specific generalization of finite symplectic groups
over modular rings.
\end{abstract}


\section{Introduction}

The Heisenberg Lie algebra and the Heisenberg-Weyl group lie at the
heart of quantum mechanics \cite{Weyl}. Therefore their symmetries
induced by unitary automorphisms play very important role in quantum
kinematics as well as quantum dynamics. The growing interest in
quantum communication science has pushed the study of quantum
systems with finite-dimensional Hilbert spaces to the forefront,
both single systems and composite systems. For them the finite
Heisenberg groups provide the basic quantum observables. It is then
clear that symmetries of finite Heisenberg groups uncover deeper
structure of finite-dimensional quantum mechanics.

Our continuing interest in finite-dimensional quantum mechanics goes
back to the paper \cite{StovTolar84} where finite-dimensional
quantum mechanics was developed as quantum mechanics on
configuration spaces given by finite sets equipped with the
structure of a finite Abelian group. In our recent paper
\cite{normalizer} detailed characterization was given of the
symmetry groups of finite Heisenberg groups for composite quantum
systems consisting of two subsystems with arbitrary dimensions $n$,
$m$. In this contribution these results for bipartite systems are
extended to the general situation where the composite systems are
multipartite, consisting of an arbitrary finite number $k$ of
subsystems with arbitrary dimensions $n_{1},\dots,n_{k}$. Their
Hilbert spaces are given by $k$-fold tensor products of Hilbert
spaces of dimensions $n_{1},\dots,n_{k}$. The exposition starts with
relevant facts of quantum mechanics. After introductory material on
finite-dimensional quantum mechanics in sections 2--3 the new
symmetry groups are described in section 4. They deserve to be
called generalized finite symplectic groups.

Symmetries in Hamiltonian mechanics belong to the class of canonical
transformations of the phase space. The symmetries considered here
have their simplest occurrence as the canonical transformations
which leave the form of fundamental Poisson brackets unchanged. In
the phase space $\mathbb{R}^{2n}$ they are the linear canonical
transformations forming the symplectic group $\Sp(2n,\mathbb{R})$
\cite{Folland}. For $n=1$ degree of freedom it reduces to
$\Sp(2,\mathbb{R})\cong \SL(2,\mathbb{R})$ with the action on the
canonical variables $q$, $p$
 \begin{equation}
        (q',p') = (q,p)\mathbb{A}=(q,p) \left( \begin{array}{cc}
        a & b \\
        c & d
       \end{array}  \right), \quad \det \mathbb{A}=ad-bc=1.
       \end{equation}

In quantum mechanics the canonical observables satisfying the
canonical commutation relations $[\hat{Q}_j,\hat{P}_k]=i\hbar
\delta_{jk}\hat{I}$ are represented by essentially self-adjoint
operators unitarily equivalent to well-known operators in the
Hilbert space $L^2(\mathbb{R}^n,d^{n}q)$ of the Schr\"odinger
representation. In the simplest case of the Hilbert space
$L^2(\mathbb{R}, dq)$ ($n=1$)
$$ \hat{Q}\psi(q)=q\psi(q), \quad
 \hat{P}\psi(q)=-i\hbar \frac{\partial\psi(q)}{\partial q}.$$
Unitary representation of linear transformations of $\hat{Q}_j$ and
$\hat{P}_k$ in the same Hilbert space $L^2(\mathbb{R}^n,d^{n}q)$ is
known as metaplectic representation \cite{Shale,Weil,Folland}. It is
a unitary irreducible representation $X(\mathbb{A})$ of the double
covering of $\Sp(2n,\mathbb{R})$ (called the metaplectic group)
which for $n=1$ reduces to the double covering of
$SL(2,\mathbb{R})$,
 \begin{equation}
(\hat{Q}',\hat{P}') =(\hat{Q},\hat{P})\mathbb{A}, \quad
\det{\mathbb{A}}=1.
       \end{equation}
The same commutators $[\hat{Q}',\hat{P}']=i\hbar \hat{I}$ obviously
lead, via the Stone--von Neumann theorem, to unitary equivalence
$$ \hat{Q}' = X(\mathbb{A})\hat{Q}X(\mathbb{A})^{-1}, \quad
  \hat{P}' = X(\mathbb{A})\hat{P}X(\mathbb{A})^{-1},
$$
also known in exponential form  for the Weyl operators
$$ X(\mathbb{A})W(s,t)X(\mathbb{A})^{-1}=W((s,t)\mathbb{A}^T).
 $$

\section{Finite-dimensional quantum mechanics}

In an $N$-dimensional Hilbert space with orthonormal basis
$\mathcal{B} = \left\lbrace\ket{0}, \ket{1}, \ldots
\ket{N-1}\right\rbrace$ unitary operators $Q_N$, $P_N$ are defined
 \cite{Weyl,Schwinger,Vourdas,Kibler} by the relations
    \begin{eqnarray*}
     Q_N \ket{j} = \omega_N^j \ket{j}, \quad  P_N \ket{j} = \ket{j-1 \pmod{N}},
    \end{eqnarray*}
where $\omega_N = \exp(2\pi i/N)$ and $j=0,1,\ldots,N-1$. If
$\mathcal{B}$ is the canonical basis of $\mathbb{C}^N$,
 the operators $P_N$ and $Q_N$ are represented by matrices
    \begin{equation}
      Q_N = \mbox{diag}\left(1,\omega_N,\omega_N^2,\cdots,\omega_N^{N-1}\right)
    \end{equation}
and
    \begin{equation}
     P_N = \left(
    \begin{array}{cccccc}
     0 & 1 & 0& \cdots & 0 & 0 \\
     0 & 0& 1&  \cdots & 0 & 0 \\
     0 & 0 & 0&\cdots & 0 & 0 \\
     \vdots &   & & \ddots &   & \\
     0 & 0 &0 & \cdots & 0 & 1 \\
     1 & 0 &0 &\cdots & 0 & 0
    \end{array} \right)
    \end{equation}
They fulfil $ P_N^N = Q_N^N = I$ and commutation relation
    \begin{equation}
     \label{qnpnomega}
    P_N Q_N = \omega_N Q_N P_N,
     \end{equation}
which is analogous to the relation for Weyl's exponential form of
Heisenberg's commutation relations.
 The finite Heisenberg group of
order $N^3$ is generated by $\omega_N$, $Q_N$ and $P_N$
     \begin{equation}
     \Pi_N = \left\lbrace \omega_N^l Q^i_N P_N^j |
     l,i,j = 0,1,2,\ldots,N-1\right\rbrace.
     \end{equation}
In a geometric interpretation \cite{StovTolar84} the cyclic group
 $\mathbb{Z}_N = \left\lbrace 0,1,\ldots N-1
 \right\rbrace $
is a configuration space for $N$-dimensional quantum mechanics.
Elements of $\mathbb{Z}_N$ label the vectors of the basis
$\mathcal{B} = \left\lbrace\ket{0}, \ket{1}, \ldots
\ket{N-1}\right\rbrace$ with the physical interpretation that
$\ket{j}$ is the normalized eigenvector of position at $j \in
\mathbb{Z}_N$. The action of $\mathbb{Z}_N $ on $\mathbb{Z}_N $ via
addition modulo $N$ is represented by unitary operators $U(k) =
P_{N}^{k}$. Their action on vectors $\ket{j}$ from basis
$\mathcal{B}$ is given by
  \begin{equation}
    U(k) \ket{j} = P_N^k \ket{j} = \ket{j - k \pmod{N}}.
  \end{equation}

\section{Finite phase space and its automorphisms}

For better understanding of symmetries for general composite systems
to be described in section 4, we summarize briefly the basic ideas
of our construction in the case of a single system with
configuration space $\mathbb{Z}_N$ \cite{HPPT02}. In this case the
finite phase space \cite{Wootters87,GHW04} (also called quantum
phase space) is the toroidal lattice $\mathcal{P}_N = \mathbb{Z}_N
\times \mathbb{Z}_N$. It is simply obtained from the finite
Heisenberg group as the quotient group $\Pi_N / Z(\Pi_N)$ where
$Z(\Pi_N)$ is the center of the finite Heisenberg group,
    $$
Z(\Pi_N) = \left\lbrace (l,0,0) |l=0,1,\ldots,N-1\right\rbrace.
    $$
The correspondence
\begin{equation}
 \Pi_N / Z(\Pi_N) \rightarrow \mathbb{Z}_N \times
\mathbb{Z}_N  :  Q^i P^j \mapsto (i,j),
 \end{equation}
is an isomorphism of Abelian groups, so the quotient group is
identified with the finite phase space with elements $(i,j) \in
\mathbb{Z}_N \times \mathbb{Z}_N$.

Since the finite Heisenberg group was introduced as a subgroup of
$\GL(N,\C)$, finite phase space $\P_N$ can equivalently be seen as
an Abelian subgroup of $\Inn(\GL(N,\C))$. For $M\in\GL(N,\C)$ let
$\Ad_{M}\in \Inn(\GL(N,\C))$ be the inner automorphism of the group
$\GL(N,\C)$
 $$\Ad_{M}(X)=MXM^{-1}\,\quad \mbox{for}\, \quad X\in \GL(N,\C).$$
Then the finite phase space $\P_N$ is isomorphic to the Abelian
group
 \begin{equation}
  \P_N \cong \{\Ad_{Q_N^i P_N^j} \vert (i,j)\in \Z_N \times \Z_N \}.
\end{equation}
In this mathematical model $\P_N$ is the Abelian subgroup of
$\Inn(\GL(n,\C))$ generated by two commuting automorphisms
$\Ad_{Q_N}$, $\Ad_{P_N}$ \cite{HPPT02}.

Consider those inner automorphisms acting on elements of $\Pi_N$
which induce permutations of phase space points, i.e. of cosets in
$\Pi_N / Z(\Pi_N)$:
     \begin{equation}
\Ad_X (\omega_N^l Q^i_N P_N^j) = X\omega_N^l Q^i_N P_N^j X^{-1},
    \end{equation}
where $X$ are unitary matrices from $\GL(N,\mathbb{C})$. These inner
automorphisms are equivalent, if they define the same transformation
of cosets,
    \begin{equation}
\Ad_Y \sim \Ad_X \quad\Leftrightarrow \quad Y Q^i P^j Y^{-1}= X Q^i
P^j X^{-1}
    \end{equation}
for all $(i,j) \in \mathbb{Z}_N \times \mathbb{Z}_N$. Modulo this
equivalence, they are elements of the normalizer of $\mathcal{P}_N$
as Abelian subgroup of $\Inn(\GL(n,\C))$.

Now the group $\P_N=\Pi_N / Z(\Pi_N)$ has two generators, the cosets
$Q$ and $P$ (or $\Ad_{Q_N}$ and $\Ad_{P_N}$, respectively). Hence,
if $\Ad_Y$ induces a permutation of cosets in $\Pi_N / Z(\Pi_N)$,
then there must exist elements $a,b,c,d \in \mathbb{Z}_N$ such that
     \begin{equation}
  Y Q Y^{-1} = Q^a P^b \quad \mbox{and} \quad Y P Y^{-1} = Q^c P^d.
     \end{equation}
It follows that to each equivalence class of inner automorphisms
$\Ad_Y$ a quadruple $(a,b,c,d)$ of elements in $\mathbb{Z}_N$ is
assigned.

\noindent \textbf{Proposition} \cite{HPPT02} \textit{For integer
$N\geq 2$ there is an isomorphism $\Phi$ between the set of
equivalence classes of inner automorphisms $\Ad_Y$ which induce
permutations of cosets and the group $\SL(2,\mathbb{Z}_N)$ of $2
\times 2$ matrices with determinant equal to $1 \, \pmod{N}$,
       \begin{equation}
        \Phi(\Ad_Y) = \left( \begin{array}{cc}
        a & b \\
        c & d
       \end{array}\right), \qquad a,b,c,d \in \mathbb{Z}_N.
       \end{equation}
The action of these automorphisms on $\Pi_N / Z(\Pi_N)$ is given by
the right action of $\SL(2,\mathbb{Z}_N)$ on elements $(i,j)$ of the
phase space $\P_N=\mathbb{Z}_N \times \mathbb{Z}_N$,
       \begin{equation}
        (i',j') =  (i,j) \left( \begin{array}{cc}
        a & b \\
        c & d
       \end{array}  \right).
       \end{equation}
}

\section{Symmetries for multipartite systems}

In our paper \cite{normalizer} we presented detailed description of
the symmetry group of the finite Heisenberg group in the case of a
bipartite quantum system consisting of two subsystems with arbitrary
dimensions $n$, $m$. The corresponding finite Heisenberg group is
embedded in $\GL(N,\mathbb{C})$, $N=nm$. Via inner automorphisms it
induces an Abelian subgroup in $\Inn(\GL(N,\mathbb{C}))$. The
normalizer of this Abelian subgroup in the group of inner
automorphisms of $\GL(N,\mathbb{C})$ contains all inner
automorphisms transforming the phase space $\P_N$ onto itself, hence
necessarily contains $\P_N$ as an Abelian semidirect factor. The
true symmetry group is then given by the quotient group of the
normalizer with respect to this Abelian subgroup.

According to the well-known rules of quantum mechanics,
finite-dimensional quantum mechanics on $\Z_n$ can be extended in a
straightforward way to arbitrary finite direct products
 $\Z_{n_1}\times\dots\times\Z_{n_k}$
as configuration spaces. The cyclic groups involved describe
independent quantum degrees of freedom. The Hilbert space of such a
composite system is the tensor product
 $$ \H_{n_1}\otimes\dots\otimes\H_{n_k}$$
of dimension $N=n_{1}\dots n_k$, where $n_{1},\dots,n_{k}\in\N$.

For such $k$-partite system, quantum phase space is an Abelian
subgroup of $\Inn(\GL(N,\mathbb{C}))$ defined by
\begin{equation}
\mathcal{P}_{(n_{1},\dots,n_{k})}=
\set{\Ad_{M_{1}\otimes\cdots\otimes M_{k}}}{M_{i}\in\Pi_{n_{i}}}.
\end{equation}
Its generating elements are the inner automorphisms
\begin{equation}
e_{j}:=\Ad_{A_{j}} \quad \textrm{for} \quad j=1,\dots,2k,
\end{equation}
 where (for $i=1,\dots,k$)
\begin{equation}
A_{2i-1}:=I_{n_{1}\cdots n_{i-1}}\otimes P_{n_{i}} \otimes
I_{n_{i+1}\cdots n_{k}},\quad
 A_{2i}:=I_{n_{1}\cdots n_{i-1}}\otimes Q_{n_{i}}
 \otimes I_{n_{i+1}\cdots n_{k}}.
 \end{equation}
 The normalizer of $\mathcal{P}_{(n_{1},\dots,n_{k})}$ in
$\Inn(\GL(n_{1}\cdots n_{k},\C))$ will be denoted
$$\mathcal{N}(\mathcal{P}_{(n_{1},\dots,n_{k})}):=
N_{\Inn(\GL(n_{1}\cdots
n_{k},\C))}(\mathcal{P}_{(n_{1},\dots,n_{k})}),$$
 We need also the normalizer of $\P_{n}$ in $\Inn(\GL(n,\C))$,
$$\mathcal{N}(\P_{n}):=N_{\Inn(\GL(n,\C))}(\P_{n}),$$
 and
$$\mathcal{N}(\P_{n_{1}})\times\cdots\times\mathcal{N}(\P_{n_{k}}):=
\set{\Ad_{M_{1}\otimes \cdots\otimes M_{k}}}{M_{i}\in
\mathcal{N}(\P_{n_{i}})}\sub\Inn(\GL(N,\C)),$$
 satisfying
$$\mathcal{N}(\P_{n_{1}})\times\cdots\times\mathcal{N}(\P_{n_{k}})
\subseteq\mathcal{N}(\mathcal{P}_{(n_{1},\dots,n_{k})}).$$

Now the symmetry group $\mathcal{H}_{[n_{1},\dots,n_{k}]}$ is
constructed in several steps. First let
$\mathcal{S}_{[n_{1},\dots,n_{k}]}$ be a set consisting of $k\times
k$ matrices $H$ of $2\times 2$ blocks
\begin{equation}
H_{ij}=\frac{n_{i}}{\gcd(n_{i},n_{j})}A_{ij}
 \end{equation}
where $A_{ij}\in\M_{2}(\Z_{n_{i}})$ for $i,j=1,\dots,k$. Then
$\mathcal{S}_{[n_{1},\dots,n_{k}]}$ is --- with usual matrix
multiplication --- a monoid. Next, for a matrix $H\in
\mathcal{S}_{[n_{1},\dots,n_{k}]}$, we define its adjoint
$H^{\ast}\in \mathcal{S}_{[n_{1},\dots,n_{k}]}$ by
\begin{equation}
(H^{\ast})_{ij}=\frac{n_{i}}{\gcd(n_{i},n_{j})}A_{ji}^{T}.
 \end{equation}
Further, we need a skew-symmetric matrix
\begin{equation}
J=\diag(J_{2},\dots,J_{2})\in\mathcal{S}_{[n_{1},\dots,n_{k}]},
\quad \mbox{where} \quad
J_{2}=\left(\begin{array}{cc}0&1\\-1&0\end{array}\right).
 \end{equation}
Then the symmetry group is defined as a finite subgroup of the
monoid $\mathcal{S}_{[n_{1},\dots,n_{k}]}$,
\begin{equation}\label{group}
\mathcal{H}_{[n_{1},\dots,n_{k}]}:=\set{H\in
\mathcal{S}_{[n_{1},\dots,n_{k}]}}{\ H^{\ast}JH= J}
\end{equation}

Our first theorem states the group isomorphism:

\noindent \textbf{Theorem 1}
\begin{equation}
\mathcal{N}(\mathcal{P}_{(n_{1},\dots,n_{k})})/
\mathcal{P}_{(n_{1},\dots,n_{k})}\cong
\mathcal{H}_{[n_{1},\dots,n_{k}]}.
\end{equation}
Our second theorem describes the generating elements of the
normalizer:

\noindent \textbf{Theorem 2} \textit{The normalizer
$\mathcal{N}(\mathcal{P}_{(n_{1},\dots,n_{k})})$ is generated by
$\mathcal{N}(\P_{n_{1}})\times\cdots\times\mathcal{N}(\P_{n_{k}})
 \quad \mbox{and} \quad  \{\Ad_{R_{ij}}\},$
  where (for $1\leq i<j\leq k$)
$$R_{ij}=I_{n_{1}\cdots n_{i-1}}\otimes\mathrm{diag}
(I_{n_{i+1}\cdots n_{j}},T_{ij},\dots,T_{ij}^{n_{i}-1}) \otimes
I_{n_{j+1}\cdots n_{k}}$$
and
$$T_{ij}=I_{n_{i+1}\cdots
n_{j-1}}\otimes Q_{n_{j}}^{\frac{n_{j}}{\gcd(n_{i},n_{j})}}.$$}

Detailed proofs of these theorems will be published elsewhere. Due
to (\ref{group}) the new groups $\mathcal{H}_{[n_{1},\dots,n_{k}]}$
represent a very specific generalization of symplectic groups over
modular rings, hence may be denoted $\Sp_{[n_{1},\dots,n_{k}]}$.
Important examples correspond to composite systems consisting of
subsystems of equal dimensions $n_{1}=\dots =n_{k}$:

\noindent \textbf{Corollary} \textit{If $n_1=\dots=n_k= n$, i.e.
$N=n^k$, the symmetry group is $\mathcal{H}_{[n,\dots,n]}\cong
\Sp(2k,\Z_n)$.}

These cases are of particular interest, since they uncover
symplectic symmetry of $k$-partite systems composed of subsystems
with the same dimensions. This circumstance was found, to our
knowledge, first in \cite{PST06} for $k=2$ under additional
assumption that $n=p$ is prime, leading to $\Sp(4,\mathrm{F}_p)$
over the field $\mathrm{F}_p$. We have generalized this result in
\cite{normalizer} to bipartite systems with arbitrary (non-prime)
$n=m$ leading to the symmetry group $\Sp(4,\mathbb{Z}_n)$ over the
modular ring $\mathbb{Z}_n$. The above corollary extends this fact
also to multipartite systems. Similar result has independently been
obtained in \cite{Han}, where symmetries of the tensored Pauli
grading of $\sl(n^k,\C)$ were investigated.

\section{Comments}

Our motivation to study symmetries of finite Heisenberg groups not
in prime or prime power dimensions as in \cite{Balian,Vourdas07,
GHW04}, but for arbitrary dimensions stems from our previous
research where we obtained results not restricted to finite fields.
Especially recall our paper \cite{TolarChadz} on Feynman's path
integral and mutually unbiased bases. Also the recent paper
\cite{VourdasBanderier} belongs to this direction, by dealing with
quantum tomography over modular rings. The papers
\cite{DigernesVV,Digernes} support our motivation, too, since they
show that finite quantum mechanics with growing odd dimensions
yields surprisingly good approximations of ordinary quantum
mechanics on the real line. This suggests a promising subject of
research to extend the results of \cite{Balian,Neuhauser} on the
metaplectic representation of $\SL(2,\mathrm{F}_p)$ from finite
fields to modular rings.

The symmetry groups $\Sp_{[n_{1},\dots,n_{k}]}$ described here can
also serve as a starting point for an alternative proof of existence
of the maximal set of mutually unbiased bases in Hilbert spaces of
prime power dimensions \cite{WoottersFields89,Bandyo}. The group
theoretical construction presented in \cite{SulcTolar07} was based
on the symmetry group $\SL(2,\mathrm{F}_p)$ of the finite Heisenberg
group for Hilbert spaces of prime dimensions. In our forthcoming
paper we shall present a new proof using the symmetry groups
$\Sp(2k,\mathrm{F}_p)$ applied to Heisenberg groups for the Hilbert
spaces of prime power dimensions.

\section*{Acknowledgements}
The first author (M.K.) was supported by the project LC 505 of
Eduard \v Cech's Center for Algebra and Geometry. The second author
(J.T.) acknowledges partial support by the Ministry of Education of
Czech Republic, projects MSM6840770039 and LC06002.

\textit{This contribution to the conference QTS 7 in Prague is
devoted to the 80th birthday of Prof. H.-D. Doebner, whose
scientific work related to symmetries and quantum theory strongly
influenced our generation.}

\end{document}